\documentclass{aa}  
\usepackage{graphicx}
\usepackage{epsfig}
\usepackage{txfonts}
\usepackage{natbib}
\usepackage{ctable}
\bibpunct{(}{)}{;}{a}{}{,}
\begin{document}
   \title{Evolution of the observed Ly$\alpha$ luminosity function from $z=6.5$ to $z=7.7$ : evidence for the epoch of reionization ?
\thanks{Based on observations collected at the European Organisation for Astronomical Research in the Southern Hemisphere (ESO), Chile, Prog-Id 181.A-0485, 181.A-0717, 60.A-9284, 084.A-0749. Based on observations obtained at the Canada-France-Hawaii Telescope (CFHT) which is operated by the National Research Council (NRC) of Canada, the Institut National des Sciences de l'Univers of the Centre National de la Recherche Scientifique of France (CNRS), and the University of Hawaii. This work is based in part on observations obtained with MegaPrime/MegaCam, a joint project of CFHT and CEA/DAPNIA and in part on data products produced at TERAPIX and the Canadian Astronomy Data Centre as part of the Canada-France-Hawaii Telescope Legacy Survey, a collaborative project of NRC and CNRS. This paper includes data gathered with the 6.5 metre Magellan Telescopes located at Las Campanas Observatory, Chile.}}

   \author{
B. Cl\'ement \inst{1,2} \and
J.-G. Cuby  \inst{1} \and
F. Courbin \inst{3} \and
A. Fontana \inst{4} \and
W. Freudling \inst{5} \and
J. Fynbo \inst{6} \and
J. Gallego \inst{7} \and
P. Hibon \inst{8} \and
J.-P. Kneib \inst{1} \and
O. Le F\`evre \inst{1} \and
C. Lidman \inst{9} \and
R. McMahon \inst{10} \and
B. Milvang-Jensen \inst{6} \and
P. Moller \inst{5} \and
A. Moorwood \inst{5}\fnmsep\thanks{Deceased. Alan Moorwood's life and scientific achievements are remembered in the September 2011 (No. 145) issue of ESO's Messenger.}
 \and
K.K. Nilsson \inst{5} \and
L. Pentericci \inst{4} \and
B. Venemans \inst{5} \and
V. Villar \inst{7} \and
J. Willis \inst{11}
}
   \offprints{B. Cl\'ement}

   \institute{
Laboratoire d'Astrophysique de Marseille, OAMP, Universit\'e Aix-Marseille \& CNRS, 38 rue Fr\'ed\'eric Joliot Curie, 13388 Marseille cedex 13, France \and
Steward Observatory, University of Arizona, 933 N. Cherry Ave, Tucson, AZ 85721, USA \and
Laboratoire d'astrophysique, \'Ecole Polytechnique F\'ed\'erale de Lausanne (EPFL), Observatoire de Sauverny, 1290 Versoix, Switzerland \and
INAF Osservatorio Astronomico di Roma, Via Frascati 33,00040 Monteporzio (RM), Italy \and
European Southern Observatory, Karl-Schwarzschild Strasse, 85748 Garching bei M\"unchen, Germany \and
Dark Cosmology Centre, Niels Bohr Institute, Copenhagen University, Juliane Maries Vej 30, 2100 Copenhagen \O, Denmark \and
Departamento de Astrof\'{\i}sica, Facultad de CC. F\'{\i}sicas, Universidad
Complutense de Madrid, E-28040 Madrid, Spain \and
School of Earth and Space Exploration, Arizona State University, Tempe, AZ 85287, USA \and
Australian Astronomical Observatory, Epping, NSW 1710, Australia \and
Institute of Astronomy, Madingley Road, Cambridge CB3 0HA, UK \and
Department of Physics and Astronomy, University of Victoria, Elliot Building, 3800 Finnerty Road, Victoria, BC, V8P 1A1, Canada
}
\date{Received May 23, 2011; accepted September 19, 2011}

  \abstract
   {}
{Ly$\alpha$ emitters (LAEs) can be detected out to very high redshifts during the epoch of reionization. The evolution of the LAE luminosity function with redshift is a direct probe of the Ly$\alpha$ transmission of the intergalactic medium (IGM), and therefore of the IGM neutral-hydrogen fraction. Measuring the Ly$\alpha$ luminosity function (LF) of Ly$\alpha$ emitters at redshift $z=7.7$ therefore allows us to constrain the ionizing state of the Universe at this redshift.}
{We observed three 7\farcm5$\times$7\farcm5 fields with the HAWK-I instrument at the VLT with a narrow band filter centred at 1.06 $\mu$m and targeting Ly$\alpha$ emitters at redshift $z \sim$ 7.7. The fields were chosen for the availability of multiwavelength data. One field is a galaxy cluster, the Bullet Cluster, which allowed us to use gravitational amplification to probe luminosities that are fainter than in the field. The two other fields are subareas of the GOODS Chandra Deep Field South and CFHTLS-D4 deep field. We selected $z=7.7$ LAE candidates from a variety of colour criteria, in particular from the absence of detection in the optical bands.}
{We do not find any LAE candidates at $z = 7.7$ in $\sim2.4\times 10^4\mathrm{Mpc^3}$ down to a narrow band AB magnitude of $\sim$ 26, which allows us to infer robust constraints on the Ly$\alpha$ LAE luminosity function at this redshift.}
{The predicted mean number of objects at $z = 6.5$, derived from somewhat different luminosity functions of Hu et al. (2010), Ouchi et al. (2010), and Kashikawa et al. (2011) are 2.5, 13.7, and 11.6, respectively. Depending on which of these luminosity functions we refer to, we exclude a scenario with no evolution from $z = 6.5$ to $z = 7.7$ at 85\% confidence without requiring a strong change in the IGM Ly$\alpha$ transmission, or at 99\% confidence with a significant quenching of the IGM Ly$\alpha$ transmission, possibly from a strong increase in the high neutral-hydrogen fraction between these two redshifts.}

   
   \titlerunning{Evolution of the observed Ly$\alpha$ LF from $z=6.5$ to $z=7.7$ : evidence for the epoch of reionization ?}
   
   \maketitle
  \date{Received .../ Accepted ...}

\section{Introduction}
Observing high-z galaxies within the first billion years of the Universe is one of the main frontiers in extragalactic astronomy. Since the discovery, less than a decade ago, of the first astrophysical object at a redshift above 6, a Ly$\alpha$ emitter at redshift 6.56 \citep{Hu2002}, spectacular progress has been made in assembling large samples of high-redshift objects. The two main techniques for finding high-redshift galaxies is to look either for strong absorption breaks in the Ly$\alpha$ forest in broad band photometry (Lyman break galaxies -- LBGs) or for a photometric excess in narrow band (NB) filters due to the Ly$\alpha$ line (Ly$\alpha$ emitters -- LAEs). In the latter case, the NB filters are usually selected to coincide with regions of low OH emission of the night sky, leading to discrete redshift values. SuprimeCam on the Subaru Telescope has revolutionized the field by enabling large samples of LAEs to be furnished at $z=5.7$ and $z=6.5$ \citep[][and references therein]{Ouchi2010,Hu2010}. The largest samples of LBGs have been recently assembled \citep{Bouwens2011} from HST observations after the successful installation of the Wide Field Camera3 (WFC3) in May 2009, but LBGs can also be found from the ground with 8-10 m telescopes equipped with efficient near infrared (NIR) cameras \citep{Castellano2010}. Quasars at high-redshift are also found using the Lyman break technique in multi-colour datasets over very wide fields. Most of the quasars at $z > 6$ have been discovered in the Sloan Digital Sky Survey \citep{Fan2006} and from a targeted programme at CFHT \citep{Willott2010}. Finally, a few gamma ray bursts (GRBs) have been discovered at very high redshift \citep[see e.g.][for an example of a GRB at redshift 8.2]{Tanvir2009}, nicely complementing the other methods by probing the faint end of the luminosity function.

Combined with observations of the cosmic microwave background (CMB), the recent discovery of large samples of objects at high redshift allows astronomers to build a comprehensive picture of the Universe during the reionization epoch when it was 500 Myr to 1 Gyr old. Polarization measurements of the CMB from WMAP \citep{Larson2011} show a large optical depth due to Thompson scattering of electrons in the early Universe, suggesting that the reionization started at $z \sim  10.5 \pm 1.2$. Conversely, the strong increase of the optical depth in the Ly$\alpha$ forest of high-redshift quasars \citep{Becker2001,Fan2006} above $\sim 8500$ \AA\ is a likely indicator that reionization was mostly complete at a redshift of about 6. How and at what pace the reionization process has taken place in the [6--10] redshift range is more difficult to establish from observations, and is still a matter of debate. A compilation of the most recent results and constraints on the neutral-hydrogen fraction of the Universe between redshifts 5 and 11 from various probes is shown in figure 23 of \citet{Ouchi2010}.

It has been proposed for a long time to use the Ly$\alpha$ transmission by the  intergalactic medium (IGM) as a probe of its ionization state during the reionization epoch \citep[see e.g.][]{Santos2004}, hence the strong emphasis recently put on Ly$\alpha$ emission of LBGs and LAEs as more and more of these objects become available. Follow-up observations of high-z LBGs at $z>6$ is now underway to detect the Ly$\alpha$ line in emission in spectroscopy. \citet{Stark2011} measure an increasing fraction of LBGs with strong Ly$\alpha$ emission from $z \sim 3$ to $z \sim 6$, and conjecture that  Ly$\alpha$ emission should remain strong at higher redshifts unless the neutral-hydrogen fraction of the IGM suddently increases. Conversely, \citet{Fontana2010} report a low fraction of Ly$\alpha$ emitters in a sample of $z > 6.5$ LBGs. These are preliminary results based on still modest spectroscopic samples, and it is expected that ongoing and new observations will clarify the situation in a near future.

Another observational method of probing the Ly$\alpha$ IGM transmission is to study the evolution of the LAE luminosity function (LF) with redshift. \citet{Ouchi2010} and \citet{Kashikawa2006,Kashikawa2011} infer from their observations that the evolution of the Ly$\alpha$ LAE LF between $z = 5.7$ and $z = 6.5$ can be attributed to a reduction of the IGM Ly$\alpha$ transmission of the order of 20\%, which can in turn  be attributed to a neutral-hydrogen fraction $x_\mathrm{HI}$ of the order of 20\% at $z = 6.5$ \citep[see e.g.][]{Ouchi2010}. Various models are elaborated to reproduce this claim, which has generated considerable interest \citep[see e.g.][]{Kobayashi2010,Dayal2011,Laursen2011,Dijkstra2010}. However, the universality of the $z=6.5$ LF from \citet{Kashikawa2006} and \citet{Ouchi2010} has recently been questioned. \citet{Hu2010} report significantly different LF parameters from the observations and analysis of a spectroscopically confirmed sample of NB selected LAEs. Similarly, \citet{Nakamura2011} report significantly lower number counts that they tentatively attribute to cosmic variance. Differences in the selection criteria and in extrapolations of the spectroscopic samples to photometric samples might partly explain the discrepancies between the various Ly$\alpha$ LAE LFs available in the literature: \citet{Kashikawa2011} have carried out extended spectroscopic confirmation of their earlier photometric sample, resulting in luminosity functions closer to the ones of \citet{Hu2010}. \citet{Cassata2011} report the results from a pure spectroscopic sample of (mostly) serendipitous Ly$\alpha$ emitters found in deep spectroscopic samples with VIMOS at the VLT ; this sample is consistent with a constant LAE luminosity function from $z \sim 2$ to $z \sim 6.6$ as reported in the literature before the recent results from \citet{Hu2010}.

The current situation at $z \ga 6$ is therefore unclear, with somewhat contradictory observational results. This hampers the validation of the reionization models and of our understanding of this key epoch of the Universe. The discrepancy limits how well we can understand reionisation during this key epoch of the Universe. New data at $z\sim6$ will help in resolving the current contention between observational results, while data at higher redshifts can bring new constraints at still poorly explored redshifts. In view of the strong interest in studying Ly$\alpha$ emission at high redshifts, searching LAEs at $z > 7$ is underway from various groups \citep{Hibon2010,Tilvi2010,Nilsson2007}. Finding $z \sim 7$ objects is not only interesting for probing the reionization epoch, but also for assessing the physical properties of these objects, which in turn allow constraining how and when they formed. Due to the extreme faintness of these very high-redshift objects, deriving their properties can only be done statistically over large samples \citep{Bouwens2011} or on individual objects that are gravitationally amplified. For instance, \citet{Richard2011} infer a redshift of formation of $18 \pm 4$ for a gravitationally amplified object at $z = 6.027$.

This paper presents new results on the Ly$\alpha$ LAE LF at $z = 7.7$, from observations carried out at the VLT with the HAWK-I instrument. This paper is organized as follows. In section~\ref{sec:observations} we describe the observations and the data reduction in section~\ref{sec:data_reduction}. In section~\ref{sec:selection} we describe our selection procedure of the $z = 7.7$ LAE candidates. In section~\ref{sec:lf} we present the constraints that we infer from our results on the $z = 7.7$ Ly$\alpha$ LAE LF, before discussing our results in section~\ref{sec:discussion}. 

We use AB magnitudes throughout this paper. We assume a flat $\Lambda$CDM model with $\Omega_{M} = 0.30$ and $H_{0} = 70$ km $\, \textnormal{s}^{-1}\, \textnormal{Mpc}^{-1}$.

\section{Observations}\label{sec:observations}
This work is primarily based on extremely deep NIR imaging data obtained with HAWK-I at the VLT, using an NB filter at 1.06 $\mu$m (hereafter referred to as \textit{NB1060}). Thanks to its wide field of view (7\farcm5$\times$7\farcm5), excellent throughput and image quality, HAWK-I is ideally suited to searching for faint NIR objects such as very high-redshift galaxies. The main data set was obtained through a dedicated ESO large programme between September 2008 and April 2010. In addition, we include in our analysis HAWK-I science verification NB data taken in 2007. We also make use of various optical and NIR broad band data, publicly available and/or from our own large programme.

\subsection{Fields}\label{sec:fields}
\begin{figure*}[hbt!]
\begin{center}
    \begin{tabular}{ccc}
      \resizebox{0.277\hsize}{!}{\includegraphics{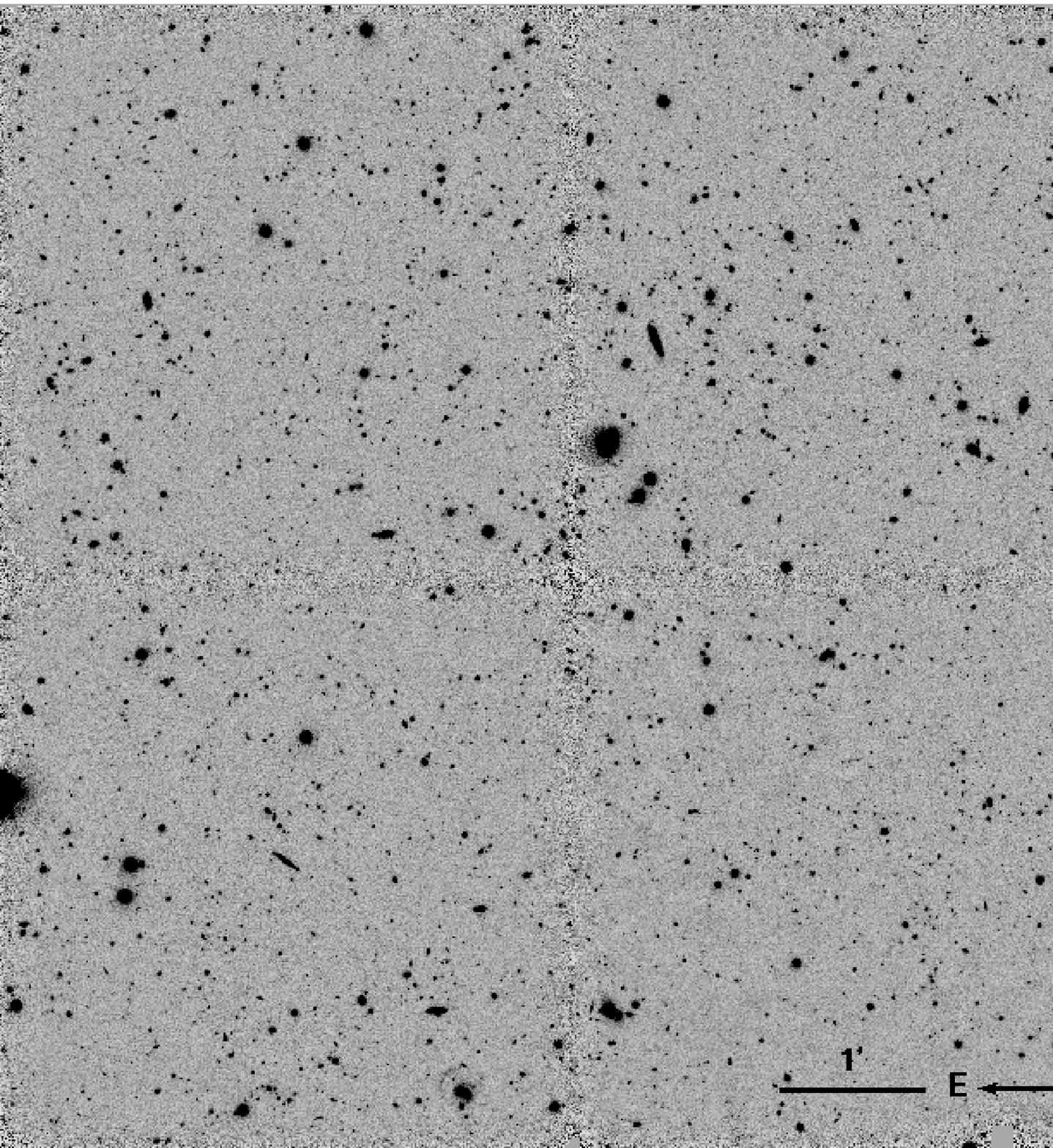}} &
      \resizebox{0.28\hsize}{!}{\includegraphics{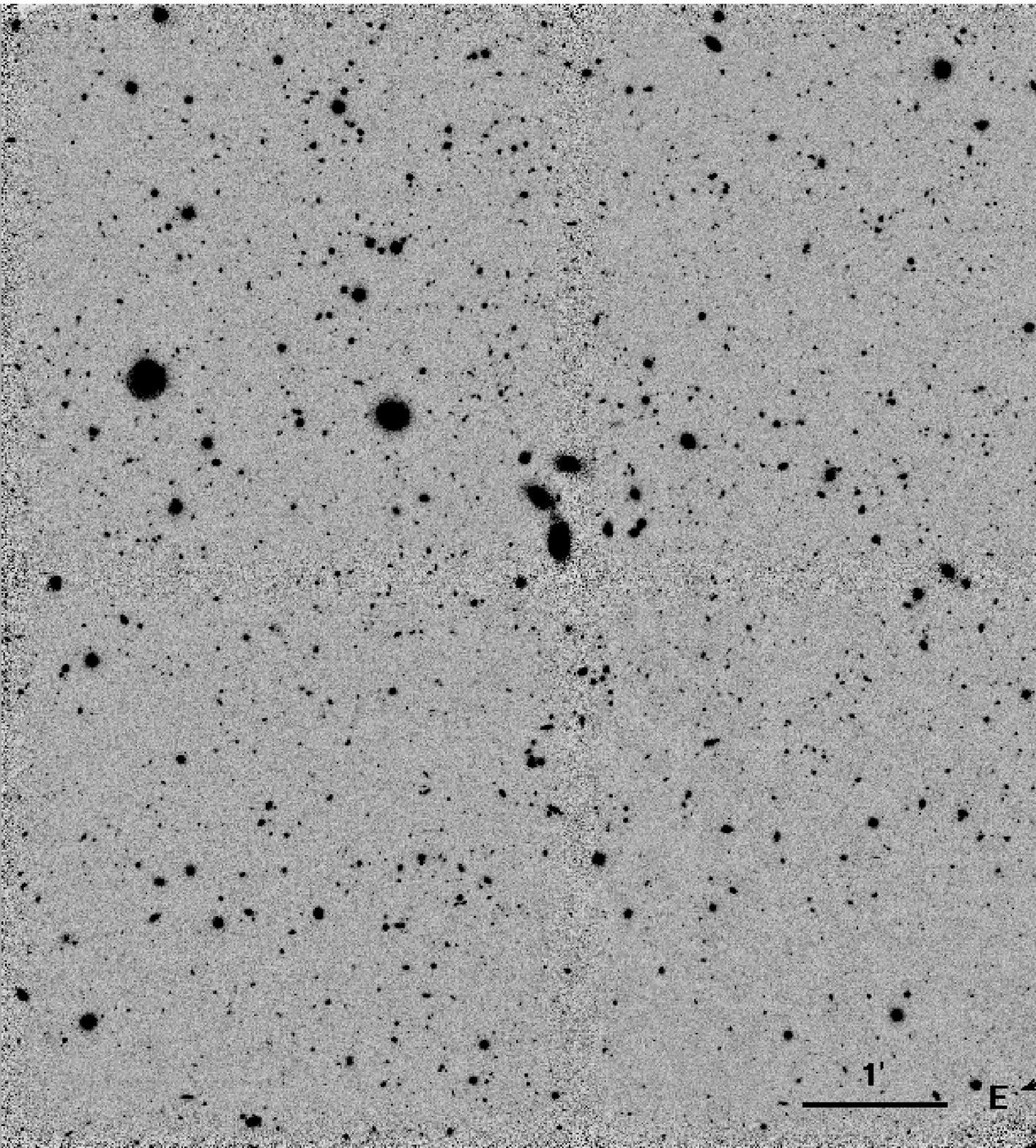}} &
      \resizebox{0.28\hsize}{!}{\includegraphics{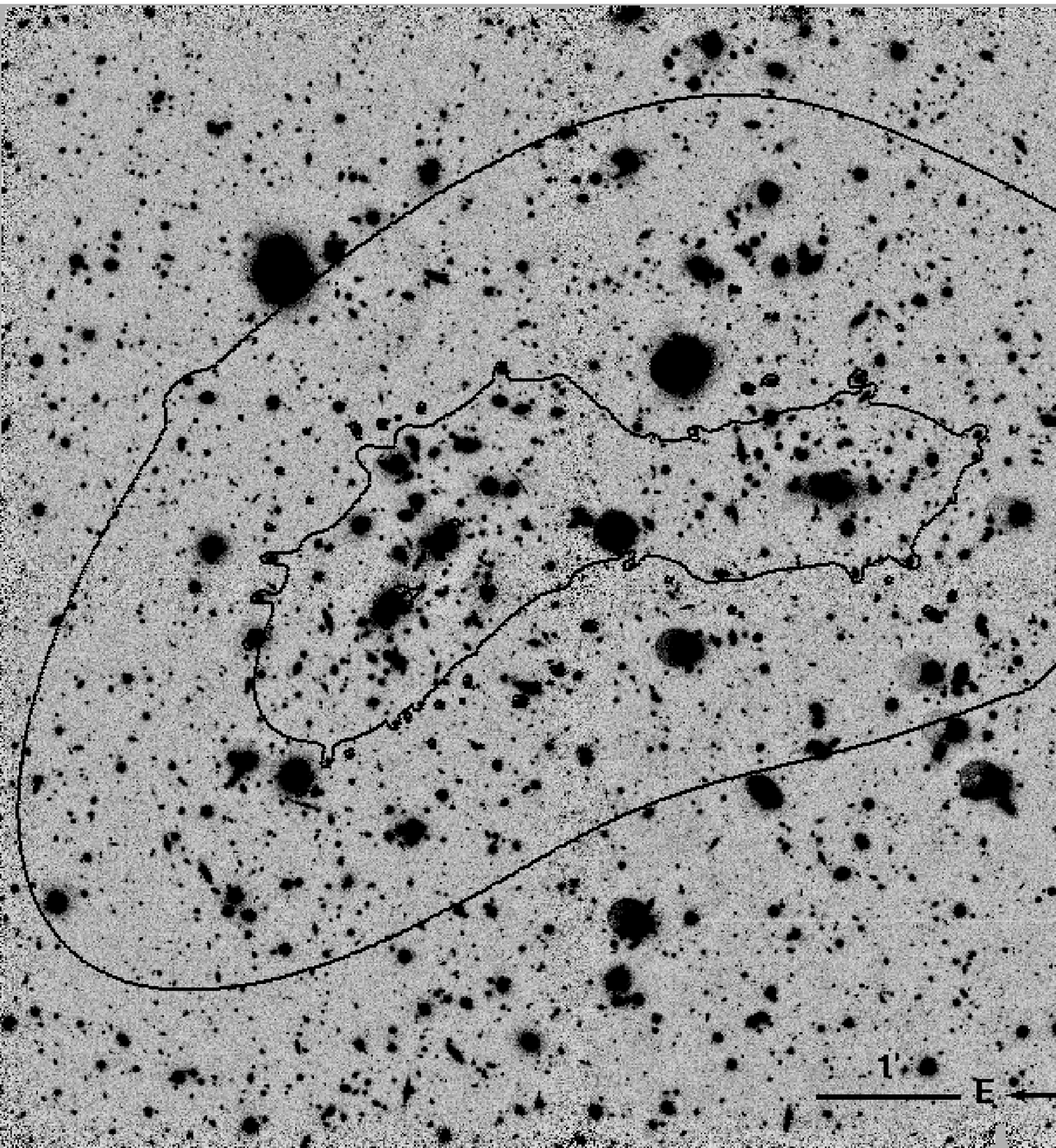}} 
    \end{tabular}
  \end{center}

\caption{Images of the CFHTLS-D4 (left), GOODS-South (centre), and Bullet Cluster (right) fields as in the final \textit{NB1060} image stacks. The inner and outer black contours on the Bullet Cluster image represent the regions where the gravitational amplification is respectively $\ge2.5$ ($\Delta m\le-1$) and $\ge1.2$ ($\Delta m\le-0.2$) for a source at redshift $z=7.7$.}
\label{fig:findingCharts}
\end{figure*}

In preparing the proposal, we carefully balanced the relative merits of blank fields and cluster fields. While gravitational amplification of background sources by foreground massive galaxy clusters allows us to probe luminosities that are intrinsically fainter than in the field, this is at the expense of areal coverage due to space distortion. The relative merits of blank and cluster fields depend on the shape of the luminosity function (LF) of the objects that are being searched, and on the properties of the observations such as field of view, integation time and overheads \citep{Maizy2010,Richard2008}. From the Ly$\alpha$ LAE LF at $z = 6.5$ that was available at the time of proposal preparation, we computed that either type of fields should yield approximately the same number of targets, while probing different (unlensed) luminosity ranges. We also analysed the balance between wide-shallow and narrow-deep survey strategies. For a total time of about 100 hrs (in the NB filter only), it was deemed that observing four fields in total would be optimal in terms of high-z LAE yield, while mitigating the effects of cosmic variance. Operational constraints, such as the distribution of the fields in right ascension, were additionally taken into account when selecting the fields. Our selected fields were Abell 1689 (13$^h$11$^m$30$^s$, -01\degr20\arcmin35\arcsec, J2000) and 1E0657-56 (Bullet Cluster) (06$^h$58$^m$29$^s$, -55\degr57\arcmin16\arcsec, J2000) for the cluster fields, the northern half of the GOODS-S field (03$^h$32$^m$29$^s$, -27\degr44\arcmin42\arcsec, J2000) and a subarea of the one square degree CFHTLS-D4 field (22$^h$16$^m$38$^s$, -17\degr35\arcmin41\arcsec, J2000) for the two blank fields.

For Abell 1689, although an extensively studied field, it proved hard to assemble a consistent multiwavelength dataset covering the full 7\farcm5$\times$7\farcm5 HAWK-I field of view. This field is therefore not included in the present analysis and it will be analysed separetely. The Bullet Cluster is a massive merging cluster that allowed the first direct empirical proof of the existence of dark matter by the combination of strong and weak-lensing analyses \citep{Clowe2006,Bradac2006}. Both clusters have well-constrained mass models and provide a lens magnification of at least a factor of 1.2 over 50\% of the HAWK-I field of view (see Figure~\ref{fig:findingCharts}). The GOODS-S and CFHTLS-D4 field were chosen for the wealth of multiwavelength data, in particular deep optical data, publicly available. For the CFHTLS-D4 field, we chose the location of the HAWK-I observations where NIR data were available\footnote{\label{note_wirds} From the CFHT WIRCam Deep Survey (WIRDS), see \texttt{http://terapix.iap.fr/rubrique.php?id\_rubrique=261}} (Bielby et al., in preparation), and paying attention to avoiding the brightest stars present in this field. Figure~\ref{fig:findingCharts} shows the finding charts corresponding to our observations inside the CFHTLS-D4, GOODS-S and Bullet Cluster fields.

\begin{figure}[bt]
\centering
\resizebox{\hsize}{!}{\includegraphics{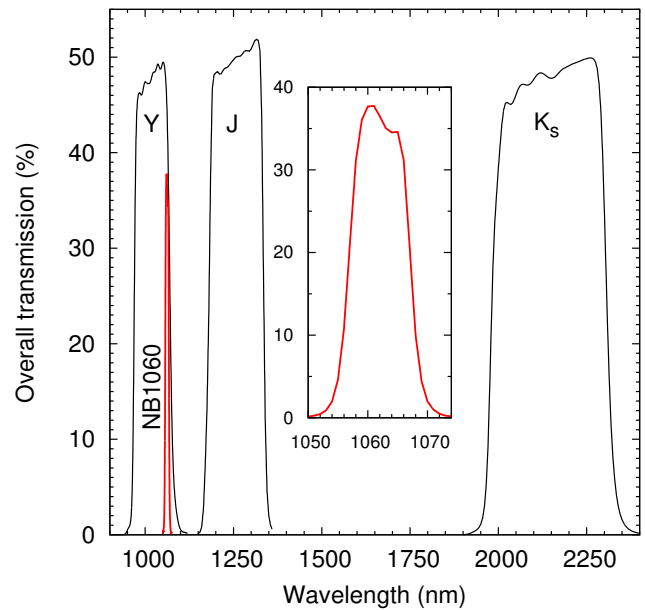}}
\caption{Transmission curves of the HAWK-I broad band and narrow band filters corresponding to the observations made as part of our large programme. The inset shows the profile of the \textit{NB1060} filter.}
\label{fig:filters}
\end{figure}

Table~\ref{tab:observations} summarizes the various observations made as part of our large programme on each of the three fields considered in the present analysis. Figure~\ref{fig:filters} shows the overall transmission curves of the HAWK-I broad band and NB filters corresponding to these observations. Table~\ref{tab:ancillary_data}  summarizes the main ancillary broad band data used in this work. Our large programme data consists of more than 110 hrs of on-sky integration time, of which $\sim$80 hrs are \textit{NB1060} data.

\begin{table*}
\caption{HAWK-I narrow band and broad band observations from our large programme}
\label{tab:observations}
\centering
\begin{tabular}{c c c c c } \hline \hline
  Field       & Filter     &    Exposure time &   Seeing &  Limiting magnitude\tablefootmark{a}\\
             &              &          (hrs)  & (\arcsec) &   \\ \hline
  CFHTLS-D4  & \textit{NB1060}   &       26.7             &    0.53       &     26.65        \\ 
  CFHTLS-D4  & \textit{J}        &         5.0          &      0.46     &      26.55       \\ 
  CFHTLS-D4  & \textit{K}$_{\mathrm{s}}$  &   0.83         &      0.50     &     24.6        \\ \hline
  GOODS-S      & \textit{NB1060}   &          31.9          &       0.58    &        26.65     \\ 
  GOODS-S      & \textit{J}        &         3.3           &     0.44      &        26.55     \\ \hline
BULLET CLUSTER& \textit{Y}        &          6.1          &      0.59     &      26.50       \\ 
BULLET CLUSTER& \textit{NB1060}   &        24.8            &      0.55     &       26.50      \\ 
BULLET CLUSTER& \textit{J}        &        6.5            &      0.49     &        26.55     \\ 
BULLET CLUSTER& \textit{K}$_{\mathrm{s}}$  &  3.75            &      0.45    &        25.45     \\ \hline
\end{tabular}
\tablefoot{
\tablefoottext{a}{$3\sigma$ aperture corrected limiting magnitude.}
}
\end{table*}

\begin{table*}
\caption{Ancillary public and private data used in this paper}
\label{tab:ancillary_data}
\centering
\begin{tabular}{c c c c c c } \hline \hline
  Field       & Filter     & Instrument    &   Seeing  &   Limiting magnitude & Reference        \\
              &            &               &   (\arcsec)         &                            &                      \\ \hline
 CFHTLS-D4   & \textit{u*}        & CFHT/Megacam       &     0.92      &    27.40\tablefootmark{a}     &1   \\ 
 CFHTLS-D4   & \textit{g'}        & CFHT/Megacam       &     0.85      &    28.20\tablefootmark{a}     &1   \\ 
 CFHTLS-D4   & \textit{r'}        & CFHT/Megacam       &     0.77      &    28.00\tablefootmark{a}     &1   \\ 
 CFHTLS-D4   & \textit{i'}        & CFHT/Megacam       &     0.73      &    27.45\tablefootmark{a}     &1   \\ 
 CFHTLS-D4   & \textit{z'}        & CFHT/Megacam       &     0.72      &    26.60\tablefootmark{a}     &1   \\ 
 CFHTLS-D4   & \textit{J}        & CFHT/WIRCam        &      0.70     &    25.10\tablefootmark{a}       & 2                 \\
 CFHTLS-D4   & \textit{H}        & CFHT/WIRCam        &      0.67     &    24.70\tablefootmark{a}       & 2                 \\
 CFHTLS-D4   & \textit{K}$_{\mathrm{s}}$ & CFHT/WIRCam        &   0.65        & 24.70\tablefootmark{a}  & 2                 \\ \hline
  GOODS-S      & \textit{B (F435W)}& HST/ACS       &     0.12      &   27.95\tablefootmark{a}        & 3     \\
  GOODS-S      & \textit{V (F606W)}& HST/ACS       &     0.11      &   28.10\tablefootmark{a}          &  3    \\
  GOODS-S      & \textit{I (F775W)}& HST/ACS       &     0.10      &    27.55\tablefootmark{a}      &  3    \\
  GOODS-S      & \textit{Z (F850LP)}& HST/ACS       &     0.10      &    27.25\tablefootmark{a}         &  3    \\
  GOODS-S      & \textit{J}        & VLT/ISAAC         &    0.50       &     26.00\tablefootmark{a}        & 4  \\
  GOODS-S      & \textit{H}        & VLT/ISAAC         &    0.53       &     25.35\tablefootmark{a}      &  4    \\
  GOODS-S      & \textit{K}$_{\mathrm{s}}$      & VLT/ISAAC         &       0.47  &  24.65\tablefootmark{a}   &4\\
  GOODS-S      & \textit{Y}        & VLT/HAWK-I        &    0.55       &      27.20\tablefootmark{a}       & 5    \\
  GOODS-S      & \textit{NB1060}   & VLT/HAWK-I        &     0.70      &      25.65\tablefootmark{a}       & 6      \\ 
  GOODS-S      & \textit{K}$_{\mathrm{s}}$  & VLT/HAWK-I        &    0.40    &   26.00\tablefootmark{a}       & 5   \\ \hline
BULLET CLUSTER&   \textit{R-Bessel}         &     Magellan/IMACS          &    0.60     &   28.00\tablefootmark{a}  & 7 \\
BULLET CLUSTER&   \textit{B (F435W)}         &     HST/ACS          &          0.10     &   27.95\tablefootmark{b}    &  8\\ 
BULLET CLUSTER&   \textit{V (F606W)}         &     HST/ACS          &          0.10     &   28.15\tablefootmark{b}    &  8\\ 
BULLET CLUSTER &   \textit{I (F775W)}         &     HST/ACS          &             0.10    &   28.55\tablefootmark{b}    & 8 \\ 
BULLET CLUSTER &   \textit{I (F814W)}         &     HST/ACS          &            0.10     &   28.30\tablefootmark{b}    &  7\\ 
BULLET CLUSTER &   \textit{Z (F850LP)}         &     HST/ACS          &           0.10    &   27.90\tablefootmark{b}    &  8\\ 
\hline
\end{tabular}
\tablefoot{
\tablefoottext{a}{$3\sigma$ aperture corrected limiting magnitude.}
\tablefoottext{b}{$3\sigma$ aperture ($\diameter=0\farcs60$) limiting magnitude.}
}
\tablebib{(1) CFHTLS T0006 release; (2) WIRDS T0002 release; (3) GOODS Version 2.0 HST ACS Imaging Data; (4) \citet{Retzlaff2010}; (5) \citet{Castellano2010}; (6) ESO Prog-Id 60.A-9284; (7) \citet{Clowe2006}; (8) \citet{Gonzalez2009}.
}
\end{table*}

%
\subsection{The HAWK-I NB1060 data}\label{sec:data}
HAWK-I is a 7\farcm5$\times$7\farcm5 NIR (0.97-2.31 $\mu$m ) imager installed on the ESO VLT UT4. It is equipped with four 2048 $\times$ 2048 pixels Hawaii-2RG detectors, separated by 15\arcsec\ wide gaps. The pixel scale is 0\farcs1065. The \textit{NB1060} filter has a central wavelength of 1062 nm, a full width at half maximum (FWHM) of $\Delta\lambda \sim$ 100 \AA, and is designed to match a region of low OH emission from the night sky. The filter width samples Ly$\alpha$ emission in the redshift range \textit{z} = [7.70 -- 7.78]. A detector integration time of 300 s is used for all the \textit{NB1060} images, ensuring background limited performance. Random telescope offsets within a box of 20\arcsec\ for the blank fields and 25\arcsec\ for the cluster fields is used for dithering. For each field, the \textit{NB1060} data are acquired over two epochs separated by one year, allowing us to discard transient objects that could be detected in a one-epoch stack, and not in the other.

The instrument had a thermal leak at the beginning of the first semester, which approximately doubled the total background in the \textit{NB1060} filter. After a technical intervention on the instrument after a few months of operations, the background returned to its nominal value, and only the observations of the CFHTLS-D4 field were affected. We were granted compensatory time that allowed us to recover the expected limiting magnitude but at the expense of unbalanced limiting magnitudes (by $\sim$ 0.5 mag) for the first and second epoch observations. For the two other fields the limiting magnitudes between the two epoch observations are within $\sim$0.15 magnitude.

In total, after image selection discarding images with poor image quality or too high background, the final stacks used in this analysis total integration times in the \textit{NB1060} filter of 26.7 hrs for the CFHTLS-D4 field, 31.9 hrs for GOODS-S (including science verification data) and 24.8 hrs for the Bullet Cluster (see Table~\ref{tab:observations}).

\subsection{Other imaging data}\label{sec:other_data}
In addition to the \textit{NB1060} data, we performed dedicated broad band HAWK-I observations within our large programme to complement, on a case-by-case basis, the broad band data that were available elsewhere (see Table~\ref{tab:observations}).

For the CFHTLS-D4 field, we had access to the very deep CFHTLS optical data and to the moderately deep NIR WIRDS data, but not as deep as our \textit{NB1060} data. We therefore took additional $J$ and $K_s$ data to improve the detection limit in these bands. For the Bullet Cluster, in the absence of well-established datasets in the optical and NIR bands, particularly over the full HAWK-I field of view, we devoted a significant fraction of the time on this field to get additional data in the $Y$, $J$, and $K_s$ bands. We note that the HAWK-I $Y$ filter bandpass includes, at its very red edge, the \textit{NB1060} filter bandpass. For this field we therefore secured a coherent and self-consistent dataset. We also used IMACS BVR images of the field obtained at the Magellan telescope \citep{Clowe2006}. For the GOODS-S field, we devoted a few hours of observations in the $J$ band filter to reach a limiting magnitude fainter than that achieved with the public ISAAC images. We also used a very deep $Y$-band image of this field from a separate HAWK-I large programme (181.A-0717) led by one of us.

\section{Data reduction}\label{sec:data_reduction}
All reduced image stacks and ancillary data products (e.g. weight maps, etc.) from our large programme are currently available upon request and will be made public through the ESO archive, as part of the Phase 3 process\footnote{http://www.eso.org/sci/observing/phase3.html}. We detail in this section the data reduction procedures that have been used to generate these high-level data products.

\subsection{Overview}
We use a mix of IRAF\footnote{\label{note_iraf}IRAF is distributed by the National Optical Astronomy Observatory, which is operated by the Association of Universities for Research in Astronomy, Inc., under cooperative agreement with the National Science Foundation.} and AstrOmatic\footnote{\label{note_astromatic}See \texttt{http://www.astromatic.net/}} routines for the data reduction, allowing us to control the reduction process step by step. All single quadrant frames from the HAWK-I mosaic array are reduced similarly and independently until the very final steps. After a first pass at the sky subtraction, all images are characterized in terms of PSF and photometric quality, allowing us to identify and remove low-quality images that would degrade the final products. Depending on the field, between 3\% and 13\% of the \textit{NB1060} images are discarded, mostly when the image quality is worse than 1\arcsec. This is the case for a fair fraction of the GOODS-S science verification images, for some of our large programme images that have been executed but not validated by the service observers, and, more rarely, for some fully validated images. A multi-pass sky subtraction is then performed, improving the quality of the masking of the objects at each pass. Images are then scaled to account for photometric variations and registered to a common reference frame. The final stacks are produced by averaging all images with a rejection algorithm using kappa-sigma clipping. The \textit{NB1060} photometric calibration is performed on unsaturated bright stars by interpolating their optical and 2MASS photometric data, following the approach described in \citet{Hibon2010}. Finally, the final four stacks for each quadrant are aligned onto a common astrometric reference.

\subsection{Data processing}\label{sec:processing}

\begin{enumerate}
 \item \textbf{Pre-processing}. Dark frames and twilight sky flats are combined into master calibration frames on a nightly basis. There are typically eight dark frames and about 30 twilight sky flatfield frames per night. No attempt is made to correct for the detector's non-linearity, which, according to the HAWK-I users manual, is below the 1\% level at 75\% of the detector saturation level. This could affect the accuracy of the photometric calibration performed on bright stars, and this is accounted for in the following.
 \item \textbf{Background subtraction}. The most delicate step in the reduction of NIR data is the sky subtraction. With dithered images, the classical way of estimating the sky at any particular pixel is by building a running sky frame for each science frame. This running sky frame is usually computed as the median of $N_{sky}$ frames around the central science frame to which it is subtracted. Some care is required, however, for this step to be optimally performed. First, the sky background varies, even in the \textit{NB1060} filter where the sky consists of a mixture of faint OH lines, sky continuum, and possibly faint thermal background leaking through the wings of the filter at wavelengths close to the detector cutoff. In addition, the background patterns have structures at low spatial frequencies that are changing with time, with the strongest changes occurring when the telescope crosses meridian. This is attributed to the rotation of the telescope pupil with time, with maximum velocity when passing the meridian. Therefore, the images used to generate a sky frame are carefully selected so as to have similar sky background patterns and to be close in time (within fewer than 15 days). The images thus selected are further zeroed to their median levels and normalized to their pixel to pixel standard deviation (with rejection of outliers). For each pixel, the median of its values in each running list of $N_{sky}$ frames is computed, with  kappa-sigma clipping for rejecting outliers. In this step, objects are masked (meaning that the values of the pixels where objects are detected are not included in the median determination) to avoid biasing the estimation of the sky toward high values. The masking process is initiated on individual sky-subtracted frames (where only the brightest objects are detected) and then repeated several times on combined stacks as described in step \# 4.

\item \textbf{Bad pixels removal}. Once the initial sky subtraction is performed on each science frame, bad pixel maps are generated on a nightly basis. Here again, we use the fact that with dithered data an object moves across the detector while bad pixels do not. Individual pixel values exceeding $\pm4\sigma$ of the local standard deviation over more than 70\% of the frames in a given night are flagged as bad pixels and replaced by a linear interpolation of the surrounding pixel values along image lines.
\item \textbf{Object masking for sky subtraction}. After the initial sky subtraction step and bad-pixel removal, the images are registered using a first-order astrometric solution and median-stacked with rejection of outliers. A mask is then generated from all objects detected in this image, together with detector regions of poor cosmetics. This mask is then used to reprocess the sky frames as described in step \#2, after which a new image stack is produced. Steps \#2 to \#4 are typically repeated three to five times until the background around the objects in the final stack is flat. This iterative procedure improves the quality of the sky subtraction, which otherwise results in overestimated sky levels noticeable as dark regions around the bright objects and in larger photometric errors. As in step \#2, the final sky frames are subtracted to the central science frames after zeroing to their median values and scaling to their standard deviations. Faint low-frequency sky subtraction residuals may still remain at this stage, which are removed with a bi-cubic-spline interpolation of a meshed background frame generated by SExtractor$^{\ref{note_astromatic}}$ \citep{Bertin1996}.
\item \textbf{Correction of photometric variations}. Frame-to-frame scaling factors are derived from the number counts measured on bright and unsaturated stars detected in each individual sky-subtracted frame. These scaling factors account for variations of the atmosphere transparency and/or of the airmass. Between two to ten stars per quadrant frame are typically used and the fluxes derived from the SExtractor MAG\_AUTO measurements. For all three fields, the variations of these frame-to-frame scaling factors are below 10\% peak-to-peak, with a star-to-star variation within each frame of about 1.5\%.
 \item \textbf{Image registration}. A relative astrometric solution is computed for each sky-subtracted frame using Scamp$^{\ref{note_astromatic}}$ and a fourth-order polynomial fit of bright star positions across the detector plane. All the resulting images are then resampled to a common reference frame with Swarp \citep{Bertin2002} using a LANCZOS4 interpolation kernel and a pixel scale of 0\farcs1065. The interpolation introduces correlated noise between pixels, and this is accounted for when computing the signal-to-noise ratio of the object as discussed in section ~\ref{sec:noise}. The accuracy of the image registration is well within one pixel for the whole data set.
\item \textbf{Final stacks and weight maps}. The final stacks for each quadrant are finally produced by averaging with 4$\sigma$ rejection the individual science frames processed as described. In this process, a map identifying the rejected pixels and a sigma map are produced. In the latter, sigma ($\sigma$) is the standard deviation of the $N$ input pixel values, excluding the rejected ones, entering into the stacks. The weight maps are then derived by computing $N/\sigma^2$.

 \item \textbf{Absolute photometric calibration}. 
The broad band \textit{J} and \textit{K}$_{\mathrm{s}}$ data taken as part of our main programme are photometrically matched to existing photometric catalogues and images of the fields in these filters. We carefully select stars in our HAWK-I images for the photometric match. For the CFHTLS-D4 field, the stellar samples consist of 67 and 58 stars in the \textit{J} and \textit{K}$_{\mathrm{s}}$ bands. The zeropoints of the HAWK-I images are adjusted to match the photometry of these samples to the photometry of the same stars in the WIRDS  data (see section~\ref{sec:fields}).  This process leaves residuals between 0.03 and 0.05 magnitude rms in the  \textit{J} and \textit{K}$_{\mathrm{s}}$  bands, respectively. The corresponding magnitudes are found to be in very good agreement with the photometry of the 19 2MASS stars present in the field, which is no surprise considering that the WIRDS data were calibrated against 2MASS. We check that there are no systematic offsets in the colours of our stellar samples compared to the colours determined from the stellar library of \citet{Pickles1998} and from a variety of stellar spectra models at various temperatures and metallicites (\citet{Marigo2008} and \texttt{http://stev.oapd.inaf.it/cgi-bin/cmd}). For the GOODS-S field, we use the publicly available ISAAC \textit{J}/\textit{H}/\textit{K}$_{\mathrm{s}}$ catalogue \citep{Retzlaff2010} to compute the zeropoint of our HAWK-I \textit{J}-band image. For the Bullet Cluster field, the \textit{J}-band and \textit{K}$_{\mathrm{s}}$-band images are calibrated from the 2MASS catalogue, leaving residual errors of 0.05 mag rms in both bands.

For the calibration of the \textit{NB1060} data, because photometric standards in narrow band filters do not exist, we perform the calibration directly on the image stacks, following the approach detailed in \citet{Hibon2010}. It consists in interpolating the \textit{NB1060} stellar photometry from the optical and NIR broad band data. This is justified by the large number of photometric datapoints available in at least two of our fields and by the absence of features at 1.06 $\mu$m in the infrared spectra of stars of spectral types earlier than M5 -- the coldest stars in our samples as determined by fitting their spectral energy distribution with the stellar models mentioned above. In practice the procedure consists in performing an ad hoc cubic spline fitting, for each star, of their magnitudes in all available bands. The magnitudes in the \textit{NB1060} band are derived from this fit. The procedure is adjusted according to the broad band data available in each field. We use exactly the same approach for calibrating the \textit{Y} image in the case of the Bullet Cluster.

For the CFHTLS-D4 field we use the \textit{u*, g', r', i', z'} optical data from the T0006 CFHTLS release and the NIR \textit{J}, \textit{H}, and \textit{K}$_{\mathrm{s}}$ WIRDS data mentioned above. The selection of stars that are neither too bright nor too faint in any of the available images leaves a sample of 23 objects. The residual error on the determination of the zeropoint from this sample is 0.05 mag rms after rejection of outliers. 

For the GOODS-S field, the optical \textit{F435W}, \textit{F606W}, \textit{F775W}, and \textit{F850LP} magnitudes are taken from the merged HST/ACS catalogue (version r2.0z) available on the GOODS website. In this field, there are 44 suitable stars, and the process leaves a residual error of 0.06 mag rms in determining the \textit{NB1060} zeropoint.

For the Bullet Cluster we use a slightly modified procedure to calibrate the \textit{Y} and \textit{NB1060} images. This is because of the lack of optical data for the entire field of view covered by HAWK-I, preventing us from performing a robust interpolation based on a large number of stars between the two wavelength ranges. Instead, we empirically determine the \textit{NB1060} zeropoints by matching the \textit{J} vs \textit{J} $-$ \textit{NB1060} (resp. \textit{K}$_{\mathrm{s}}$ vs \textit{NB1060} $-$ \textit{K}$_{\mathrm{s}}$) colour-magnitude diagrams of the stars present in this field to the same diagrams produced on the GOODS-S and CFHTLS-D4 fields after calibration. The same procedure is used to calibrate the \textit{Y}-band image by comparing it to the colours of the stars in the GOODS-S field for which \textit{Y} band data are available.

All the procedures described above are conducted quadrant by quadrant, with a further iteration on the full four quadrant images. In total, considering the consistency between the many checks that are performed, and despite the various methods used, we estimate that the final accuracy of the photometric calibration is of the order of 0.1 magnitude rms.

 \item \textbf{Absolute astrometric registration and final image stitching}. The last step in our reduction process consists in stitching and registering the four detector images to the reference images of each field. The final astrometric solution is computed by Scamp$^{\ref{note_astromatic}}$ with a fourth-order polynomial fit of the star positions. The CFHTLS-D4 stacks are aligned to the archival CFHTLS images. The GOODS-S stacks are aligned to the optical HST/ACS images, and the Bullet Cluster stacks are aligned to the 2MASS catalogue in the absence of astrometrically calibrated data across the entire area. The final astrometric residuals are below the 0\farcs05 rms level across the entire field of view for all images. Finally, the resampling and final image stitching are performed using Swarp \citep{Bertin2002}. In addition, Swarp propagates the astrometric solution to the weight maps and uses a weighted mean to compute pixel values in the small overlap between quadrants due to the dithering pattern.
\end{enumerate}

\subsection{Final image properties}\label{sec:properties}
We now discuss the global properties of the final images: image quality (\textit{FWHM}), noise, and detection limits. For consistency, the same procedure is applied to all the images used in this work, including archival data. The \textit{FWHM} and the detection limits are listed in Table \ref{tab:observations}.

\subsubsection{Image quality}\label{sec:iq}
\label{sec:IQ}
The image quality is determined from a high signal-to-noise ratio point spread function (PSF) generated by stacking unsaturated and isolated stellar images (range of 20--30) in each field. After normalization, the stellar images are centred and median-stacked with a 4$\sigma$ outlier rejection. The rejection reduces the contribution from faint neighbouring objects in the wings of the PSF but does not affect its profile. The resulting \textit{NB1060} images are slightly elongated, for reasons that are unknown to us, with a measured ellipticity from about 0.05 to 0.1 along a direction $\le 10^\circ$ away from the N-S axis. The \textit{FWHM} values are derived from a 2D-Gaussian fit to the median profile. The three \textit{NB1060} final images have exquisite image qualities ranging from 0\farcs53 to 0\farcs58 (see Table~\ref{tab:observations}).

\subsubsection{Photometric errors and correlated noise}\label{sec:photometry}
\label{sec:noise}
Image resampling introduced by the distortion correction, shifting, stitching, and registration processes introduces correlation in the noise of the images. This leads in turn to underestimating the photometric errors when considering the pixel-to-pixel noise properties, see e.g. \citet{Grazian2006}, Appendix A, or \citet{Casertano2000}. To measure and account for this well known effect, we carefully analyse the noise properties of the images over apertures of varying sizes and derive correction factors that we can then apply to the photometric data measured by SExtractor. Indeed, SExtractor derives the photometric error for each object it finds by computing the local pixel-to-pixel noise fluctuation in the vicinity of the object (in the faint-object, background-limited regime). The SExtractor photometric errors are therefore affected by the correlation of the noise. For each image, we select a thousand positions corresponding to source-free background regions determined from the final SExtractor segmentation (object mask) image. For each position, we measure the integrated flux in circular apertures of diameters ranging from $N_{ap}$ = 1 to 25 pixels (0\farcs1065\ to 2\farcs663). For a given aperture size, the variance of these fluxes, $\Delta F_{ap}^2$, differs, because of the correlated noise, from the variance $N_{ap} \times \sigma_1^2$ of the errors computed from the pixel-to-pixel variance $\sigma_1^2$ and the number of pixels $N_{ap}$ in the aperture. The ratio of these two quantities $f_{corr}=(\Delta F_{ap}^2 / (N_{ap} \times \sigma_1^2))$ gives the noise-correction factor that can be used to correct the photometric errors measured by SExtractor. The ratio $f_{corr}$ clearly depends on the aperture size: for apertures smaller than the correlation length of the noise, which is related to the size of the resampling interpolation kernels, $f_{corr} \sim 1$, whereas for large apertures $f_{corr} \propto N_{pix}^{0.1}$, ranging from 1.11 for a three-pixel diameter aperture to 1.67 for a 25 pixel diameter aperture. The procedure is repeated ten times for each field, and all three fields give similar and consistent measurements.

This analysis finally allows us to assign signal-to-noise ratios  (\textit{SNR}) to the objects detected by SExtractor, in the sky background limited regime, using
\begin{equation}
SNR = \frac{F}{f_{corr}\,\Delta F_{SE}}
\end{equation}
where $\Delta F_{SE}$ is the photometric error measured by SExtractor. The relation between \textit{SNR} and the magnitude error $\Delta m$ is finally given by

\begin{equation}
\Delta m = 1.086 / SNR.
\end{equation}

\subsubsection{Aperture corrections and optimal apertures}\label{sec:aperture_correction}
\label{sec:aper_corr}
We measured curves of growth on unsaturated stars for all image stacks used in this work, using apertures between 1 and 150 pixels (0\farcs1065 to 16\arcsec) in diameter. Less than 1\% of the flux resides in the wings of the PSF beyond radii of 7\arcsec, and we therefore safely use the 16\arcsec aperture correction to estimate the total flux of unresolved or moderately resolved objects. From the curves of growth of both flux and noise, we derived the optimal diameter that maximizes the signal-to-noise ratio for point-like objects. In practice, for all of our \textit{NB1060} image stacks, a diameter of $\diameter=0\farcs64$ (6 pixels) is used. The corresponding aperture corrections $\delta m_{ap}$ for the three \textit{NB1060} final images are $\delta m_{ap}=0.90\pm0.04$ mag, $\delta m_{ap}=0.82\pm0.03$ mag, $\delta m_{ap}=0.85\pm0.04$ mag for the GOODS-S, CFHTLS-D4, and Bullet Cluster fields, respectively. The corresponding noise correcting factors are $f_{corr}$ = 1.22, 1.14, and 1.14 for the GOODS-S, CFHTLS-D4, and Bullet Cluster fields, respectively. 

\subsubsection{Detection limits}\label{sec:det_limits}
\label{sec:limiting_magnitude}
Finally, from the two parameters defined above, $f_{corr}$ and $\delta m_{ap}$, one can define the $1\sigma$ limiting magnitude $m_{1\sigma}$ for point-like objects:

\begin{equation}
m_{1\sigma}=-2.5\log_\mathrm{10}(f_{corr} \Delta F_{SE})-\delta m_{ap}+ZP,
\end{equation}
where $f_{corr}$, $\Delta F_{SE}$ and $\delta m_{ap}$ correspond to apertures of 0\farcs64 in diameter. This leads to a $3\sigma$ \textit{NB1060} point source detection limit of $m_{3\sigma}=26.65$, 26.65, and 26.50 for the GOODS-S, CFHTLS-D4, and Bullet Cluster, respectively, as reported in Table~\ref{tab:observations}.

\section{Candidate selection}\label{sec:selection}

\subsection{Detection completeness}
\label{sec:completeness}
\begin{figure}[bt]
\centering
\resizebox{\hsize}{!}{\includegraphics{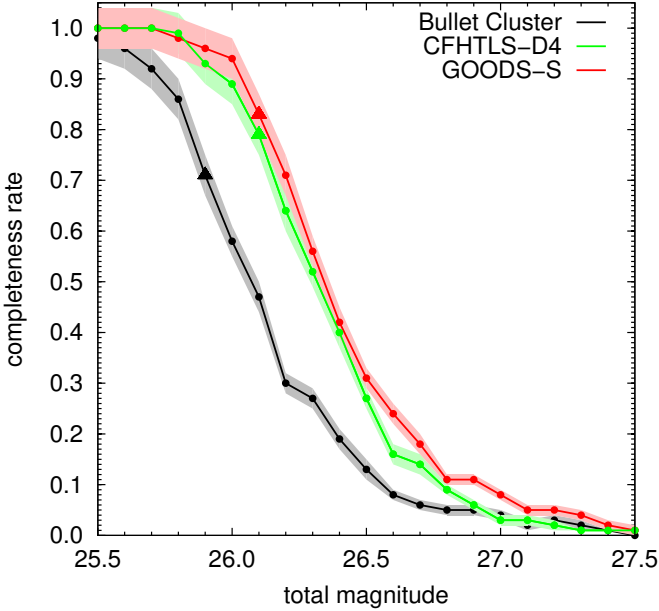}}
\caption{Completeness levels of the \textit{NB1060} images for the three observed fields. The coloured areas correspond to the Poissonian errors on the number counts in the corresponding magnitude bin. The triangles indicate the 5$\sigma$ magnitude limits. 
}
\label{fig:completeness}
\end{figure}

We use SExtractor \citep{Bertin1996} version 2.8.6 for source detection and photometric measurements. We fit the PSF median profile discussed in section ~\ref{sec:IQ} with a sum of three 2-dimensional Gaussians from which we derive the filter ($9\times9$ pixels or $0\farcs96 \times 0\farcs96$) used by SExtractor for spatial filtering during the detection process. We then use this point-source model to generate mock sources injected into the image to estimate the image detection completeness. We inject 500 mock sources per $\Delta m=0.1$ magnitude bins in regions of the images randomly distributed and free of objects. We perform a number of tests to determine the optimum SExtractor detection parameters that maximize the number of detected objects while minimizing the number of false alarms. Considering the absence of candidate LAEs in our images, we choose to push to the faintest possible limits. False alarms are investigated by running SExtractor on the negative images and are discussed in section~\ref{sec:selection_field2field}. We determine that an adequate set of SExtractor parameters, for the purpose of our analysis, is to trigger a detection on one pixel (DETECT\_MINAREA) after spatial filtering at 0.7$\sigma$ above the local background (DETECT\_THRESH). With these detection parameters, the average signal-to-noise ratio of sources in the 50\% completeness magnitude bin is of about 4, and a \textit{SNR} of 5 corresponds to a completeness rate of 70 to 80\%. Figure~\ref{fig:completeness} shows the completeness rates achieved in the three \textit{NB1060} images. 

\subsection{Selection criteria}
\label{sec:selection_criteria}
We do not expect $z\sim7.7$ LAEs to be detected in any of the filters blueward of the Ly$\alpha$ line redshifted to 1.06 $\mu$m. First, negligible amounts of radiation are expected to escape the galaxy and to be transmitted by the IGM below the Lyman limit, which is redshifted to $\sim$ 790 nm. In addition, all the radiation between the Ly$\alpha$ and Ly$\gamma$ lines at $z = 7.7$ is entirely redshifted beyond the Gunn-Peterson trough at $\sim$850 nm observed in the spectra of high-redshift quasars \citep{Fan2006}. Only in the blue part -- the most depressed part -- of the Ly$\alpha$ forest, just above the Lyman limit, can we therefore expect some flux from a $z \sim 7.7$ LAE to arrive on Earth, in the wavelength range [790--850] nm approximately. In practice, considering the limiting magnitudes of our optical and \textit{NB1060} images, an absorption of 2 magnitudes or so will result in no detections in any of the optical bands. 

For the purpose of the analysis described in this paper, we built a master catalogue of all the \textit{NB1060} detected objects, measuring their magnitudes in each of the optical and NIR broad band images by running SExtractor in double image mode. To do so, we resample all images to the HAWK-I pixel scale of  0\farcs1065. We then search objects in this master catalogue that are not detected in the optical images at an initial $SNR\le3$ level, reduced to $SNR\le2$ after visual inspection. In the case of the CFHTLS-D4 field, we request in addition that these objects not be detected at a $2\sigma$ significance level in a $\chi^2$ image of the field obtained by combining the \textit{g'}, \textit{r'}, and \textit{i'} images. In the case of the GOODS-S field, we use a similar non-detection limit on a \textit{bviz} $\chi^2$ image obtained by combining the four broad band HST/ACS images. Because the \textit{NB1060} bandpass is located within the bandpass of the \textit{Y} filter (at its red edge), the Ly$\alpha$ line may be detected in the \textit{Y} filter. To estimate the \textit{Y}$-$ \textit{NB1060} colour as a function of redshift, we generate simple synthetic models of LAE spectra, consisting of a narrow Ly$\alpha$ line and a UV-continuum of energy distribution $f_\lambda \propto \lambda^\beta$. We allow the UV slope $\beta$ to vary from -3 to 0, and we set the flux density below the Ly$\alpha$ line to zero. For objects with redshifts in the interval [$7.70-7.78$] corresponding to the \textit{NB1060} filter, we find that the \textit{Y}$-$ \textit{NB1060} colour is $\sim 2.35\pm0.35$, with the lowest values corresponding to situations where the Ly$\alpha$ line is redshifted near the edges of the \textit{NB1060} filter transmission curve, and therefore strongly attenuated. This corresponds to the case of high-redshift LBGs detected through their continuum in the \textit{NB1060} filter. As we see in the next section, the \textit{Y} images (when available) are not deep enough, relative to the \textit{NB1060} images, to measure such a colour on the faintest objects, but they do allow us, conversely, to discard blue and moderately bright objects that pass the other colour selection criteria. To secure the presence of an emission line in the \textit{NB1060} filter, we further require a 1$\sigma$ NB excess over the flux measured in the \textit{J}-band. From equation (6) in \citet{Hibon2010}, \textit{NB1060} $-$ \textit{J} $\le0$ corresponds to equivalent widths \textit{EW}$_{\mathrm{obs}}\ge 50 $\AA\,or \textit{EW}$_{\mathrm{rest}}\ge 5.7$\AA, assuming a flat continuum spectrum ($f_\nu = const$). We note that the \textit{NB1060} filter is placed approximately at the centre of the bandpass covering the  \textit{Y} and \textit{J} filters, allowing us to further constrain the presence of an LAE from its colour between the  \textit{NB1060} and  \textit{Y+J} bandpasses; however, in the absence of candidates from the criteria used so far (see next section), this did not prove necessary to add to our selection criteria. Finally, we restrict the analysis to sources having $SNR\ge5$ in the \textit{NB1060} final images and $SNR\ge2$ in partial or intermediate image stacks corresponding to different observing epochs. In summary, our detection criteria are
\begin{enumerate}
\item \textit{NB1060} $\ge5\sigma$ $\land$ \textit{NB1060}$_{epoch1}$ $\ge2\sigma$ $\land$ \textit{NB1060}$_{epoch2}$ $\ge2\sigma$,
\item no detection above the $2\sigma$ level in any of the visible broad band filters,
\item $2 \le $ \textit{Y} $-$ \textit{NB1060} $\le2.7$ (when \textit{Y} band data are available),
\item \textit{NB1060} $-$ \textit{J} $\le 0$ with $1\sigma$ significance.
\end{enumerate}

We finally note that astrophysical sources such as transients, extremely red objects (EROS), high-EW low-\textit{z} line emitters or T-dwarfs can potentially satisfy the optical non-detection criteria defined here. Relatively deep NIR \textit{Y} and/or \textit{J} and/or \textit{K}$_{\mathrm{s}}$ band data are therefore required for consolidating the selection (e.g. criteria \#3 and \#4) and reducing contamination from astrophysical sources. The \textit{J} and $\textit{K}_{\mathrm{s}}$ band data in particular are useful to identify T-dwarfs and EROs, even if the latter are often detected in deep optical images. In the absence of LAE candidate in our data, we are clearly not affected by contamination, thanks to the coherent datasets that we use. We refer the interested reader to section 3.3 of \citet{Hibon2010} for a somewhat more detailed analysis of contamination effects in a similar $z=7.7$ LAE search, in particular by $\textrm{H}\alpha$, [OIII], and [OII] line emitters.

\subsection{Selection field by field}\label{sec:selection_field2field}
\label{sec:photometric_sample}
The  5$\sigma$ detection limit in the \textit{NB1060} filter corresponds to magnitudes $m_{5\sigma}=25.9$ to $m_{5\sigma}=26.1$ depending on field (see Table~\ref{tab:observations}), and to a completeness level of 70 to 80\% (see figure~\ref{fig:completeness}). The corresponding colour criteria used for the selection of candidates differ among the three fields depending on the depth of the optical images available in each of them.

\textbf{CFHTLS-D4}. The selection criterion \#2 in the previous section corresponds to the following colour criteria:
\begin{equation}
\begin{array}{lcl}
\textit{u*}_{2\sigma}-\textit{NB1060} & \ge & 1.7,\\
\textit{g\arcmin}_{2\sigma}-\textit{NB1060} & \ge & 2.5,\\
\textit{r\arcmin}_{2\sigma}-\textit{NB1060} & \ge & 2.3,\\
\textit{i\arcmin}_{2\sigma}-\textit{NB1060} & \ge & 1.8,\\
\textit{z\arcmin}_{2\sigma}-\textit{NB1060} & \ge & 0.9.\\
\end{array}
\end{equation}

There are $\sim$ 6500 \textit{NB1060} objects detected in this field. The application of criteria \#1 and \#2 of section~\ref{sec:selection_criteria} yields 20 objects. Ten are visually identified as instrumental artefacts caused by electronic crosstalk (see section~\ref{sec:radioactivity}). Seven of the remaining objects are detected and relatively bright in the \textit{J} band and therefore rejected after application of criterion \#4. We note in passing that the brightest of these objects has $\textit{NB1060}=24.15$, $\textit{J}=23.85$ and $\textit{J}-\textit{H}_{2\sigma}\le -0.85$ and is very likely a T-dwarf. 

Amongst the three remaining objects, one is located near the edges of the image and appears sharper than the PSF. One is located in the wings (4\arcsec) of a bright extended galaxy, therefore of suspicious photometry and therefore unusable. Finally, the last one appears to be a variable, extended object, with $\ge1$ magnitude difference between the first- and second-epoch observations. All three objects are therefore discarded.

To allow for possibly slightly extended LAE candidates and for consistency checks, we carry out a second selection using larger apertures ($\diameter=1\farcs065=10\,\mathrm{pixels}\sim2\times\textit{FWHM}$), and applying the appropriate aperture correction. Criteria \#1 and \#2 yielded 14 detections, out of which nine are in common with the previous sample of 20 objects. Amongst the remaining five new detections, one is an obvious artefact near a bright star, two are detected in the \textit{J}-band and are rejected due to the low significance of their NB excess. The two last ones are low significance detections in the \textit{NB1060} image ($SNR\le5.5$) and both show extended and dubious morphologies.

Checking the robustness of the rejections further, we investigate the false alarms on the negative image using the two aperture diameters mentioned above. After removing the well-determined crosstalk features, which have a negative component, we are left with eight objects using the $\diameter=0\farcs64$ aperture and 2 using the $\diameter=1\farcs065$ aperture, all of them at the limit of our signal-to-noise ratio selection. These genuine noise artefacts have very similar morphologies to those of the positive detections that were rejected on the science image: either very sharp or extended and irregular with bright non-contiguous pixels. This legitimizes our somewhat ad hoc, but pragmatic earlier selection based on the morphology of the faintest positive candidates, and we therefore conclude that there are no $z=7.7$ LAE candidate in the CFHTLS-D4 field.\\

\textbf{GOODS-S}. 
Here the selection criterion \#2 corresponds to the following colour criteria:
\begin{equation}
\begin{array}{lcl}
\textit{F435W}_{2\sigma}-\textit{NB1060} & \ge & 2.3,\\
\textit{F606W}_{2\sigma}-\textit{NB1060} & \ge & 2.5,\\
\textit{F775W}_{2\sigma}-\textit{NB1060} & \ge & 1.9,\\
\textit{F850LP}_{2\sigma}-\textit{NB1060} & \ge & 1.6.\\
\end{array}
\end{equation}

There are $\sim$ 5100 \textit{NB1060} detected objects. The application of criteria \#1 and \#2 yields 16 objects, of which 12 are visually identified as instrumental artefacts caused by electronic crosstalk. Amongst the four remaining objects, one is due to a mismatch on a blended object, and two are marginally detected objects at the edges of the image. The last object is detected in the \textit{Y}-band but does not satisfy criterion \#3 above; interestingly, it is identified with reference G2\_1408 as a $z\sim$ 7 object in \citet[][and references therein]{Castellano2010}. This object has been followed up in spectroscopy \citep{Fontana2010}, yielding a tentative detection of the Ly$\alpha$ line at a redshift $z=6.97$. This object therefore appears to be an LBG, caught by its strong UV continuum emission detected in the \textit{NB1060} filter.

Similar to what was done on the CFHTLS-D4 field, we then performed a second selection using $\diameter=2\times\textit{FWHM}=1\farcs065$ apertures. This yields two new detections (beyond the obvious electronic artefacts): one is detected in the \textit{Y}-band and does not pass criterion \#3, and is also marginally noticeable in the optical bands. The other one has a dubious morphology and is close to a bright object, and is therefore rejected. Finally, we carried out the false alarm analysis on the negative image, and we detect a handful of events with dubious morphologies, leading to the same conclusions as for the CFHTLS-D4 field.\\

\textbf{Bullet Cluster}.
The dataset for the Bullet Cluster field is somewhat different than for the other fields. As explained in section~\ref{sec:other_data}, we accomodated a consistent set of HAWK-I data in the \textit{Y}, \textit{NB1060}, \textit{J} and \textit{K}$_{\mathrm{s}}$ bands within our large programme. In addition, we used HST/ACS images of the inner part of the field, as well as moderately deep IMACS images from the Magellan telescope \citep{Clowe2006}. There are $\sim$ 7000 \textit{NB1060} objects satisfying criterion \#1, the vast majority of which are detected in the \textit{Y} filter and do not satisfy criterion \#3. Only 127 objects are not detected in the \textit{Y}  image at 3$\sigma$, but because $\textit{Y}_{3\sigma} - \textit{NB1060}_{5\sigma} = 0.4$, it is impossible to conclude whether they satisfy criterion \#3 or not. After visual inspection and rejection of electronical ghosts and obvious artefacts, all but one object show a clear counterpart in either one of the HST/ACS images or in the IMACS images. This object shows no NB excess ($\textit{NB1060}-\textit{J}=0.55\pm0.25$), and together with a marginal detection in the \textit{Y}-band ($\textit{SNR}\sim2.4$, $\textit{Y}=26.40\pm0.45$) and a non-detection in $\textit{K}_{\mathrm{s}}$ ($\textit{K}_{\mathrm{s}}\le25.3$, $3\sigma$ upper limit), we conjecture that this object is probably a T-dwarf. Finally, a selection based on larger apertures as for the two other fields yields no new candidates. We therefore conclude, again, that there are no $z=7.7$ LAE candidates in the Bullet Cluster field.

\subsection{Instrumental artefacts}
\label{sec:radioactivity}
Instrumental artefacts are a potentially important source of contamination. As explained in section~\ref{sec:selection_field2field}, candidates are found that are rejected as instrumental artefacts. We describe here some of the instrumental artefact sources that are observed in the HAWK-I data.
\begin{enumerate}
\item Electronic crosstalk. The HAWK-I data suffer from inter-channel crosstalk from the readout electronics. This results in donut-shaped artefacts at regularly spaced intervals along detector lines where bright stars are present. These artefacts were largely attenuated after a technical intervention in the instrument that took place in May 2009. The crosstalk pattern follows the dithering pattern and is therefore present on the final stacked images. The crosstalk artefacts are easy to recognize from their shapes and fixed distances from bright stars along detector rows. Because they do not have counterparts in optical images, these artefacts are selected as candidates in our analysis, but are easily dealt with a posteriori. No attempt was made to remove these artefacts during data processing.
 \item Optical ghosts. Reflections inside the instrument generate typical out-of-focus and decentred pupil images around bright stars. The surface brightness of these haloes was measured to be $10^{-4}$ of the peak intensity in the PSF profiles. Only focussed optical ghosts can be mistaken as candidates, and no such artefacts are observed on the HAWK-I images.
 \item Persistence. Persistence from previously observed bright stars is at a fixed detector position. The persistence features therefore do not follow the dithering pattern and are rejected when combining the images with sigma clipping in the final stacks. Considering the large number of frames used in the stacks (more than 200 frames), persistence effects are unlikely to leave residuals that can be mistaken as candidates.
 \item Radioactive events. One of the HAWK-I arrays (chip \#2, Id: ESO-Hawaii2RG-chip78) suffers from a strong radioactive event rate, coming from the detector substrate \citep{Finger2008}. These radioactive events generate showers of variable intensity (typically thousands of electrons) and extent (typically a few tens of pixels on a side). Some events can be as bright as a few hundred thousand electrons and extend up to 400 pixels in one direction. Because of the long detector integration times (DIT) used for the  \textit{NB1060} images (300 seconds), there are a few tens of such events in a single frame. This results in poor background subtraction and moderately high-frequency residuals (a few tens of pixels) in individual sky-subtracted frames. Although the global noise properties are not significantly different in this quadrant than in the others, the overall cosmetics are somewhat poorer. As a consequence, most of the low signal-to-noise ratio detections and with dubious morphologies reported in section~\ref{sec:photometric_sample} appear to be in this quadrant and are therefore rejected as artefacts after visual inspection.
 \item Noise. As explained earlier, we chose a low detection threshold in order to push to the faintest detection limits, triggering a handful of low signal-to-noise ratio false alarms, particularly on the highly radioactive quadrant. We settle on a detection threshold that allows us to handle these false alarms.
\end{enumerate}

\section{Constraints on the $z=7.7$ Ly$\alpha$ luminosity function}
\label{sec:lf}
\subsection{Comoving volume}\label{sec:volume}

The effective field of view in the transverse dimension of the \textit{NB1060} stacks is computed as a function of the detection limit for each field from the background noise maps generated by SExtractor, and accounting for the correlation of the noise as described in section~\ref{sec:noise}. The sensitivity is reduced at the edges of each quadrant image thanks to the dithering process and in regions close to bright objects. All objects are masked, reducing the total effective area by $\sim$ 10\% for the blank fields, and $\sim$ 25\% for the  Bullet Cluster due to the large number of bright galaxy cluster members. For this field, we further use a map of the gravitational amplification and space distortion from a detailed model of the cluster (Richard et al. in preparation) to compute the effective, unlensed, comoving area corresponding to the image \citep[see][for a similar example]{Willis2008}. In the direction of the line of sight, we use the filter transmission curve to determine the effective width. We then approximate the \textit{NB1060} filter with a rectangular filter of width equal to this effective width. The effects of the filter transmission curve on LAE detection and comoving volumes are detailed in e.g. \citet{Willis2005} and \citet{Hu2010}. Considering our null results, these effects will not affect our conclusions. We also assume that the filter transmission curve does not vary significantly over the instrument's field of view, an assumption motivated by the relative uniformity of the sky background over the instrument's field of view. We finally assume that the observed Ly$\alpha$ emission lines are significantly narrower than the filter width, as is the case for high-z LAEs, and we therefore ignore the effect of the line width on the line flux measured through the \textit{NB1060} filter. The effective width of the \textit{NB1060} ($\Delta\lambda_{\mathrm{eff}} \sim \mathrm{100}$\AA) defines the [$7.70-7.78$] redshift interval probed by our observations.

Converting  magnitudes to line luminosities requires assumptions on the equivalent width (EW) of the Ly$\alpha$ line. Distributions of observed Ly$\alpha$ lines in high-z LAEs vary from a few tens of Angstroms to lower limits of a few hundred Angstroms \citep[see e.g.][]{Taniguchi,Ouchi2010}. The \citet{Taniguchi} distribution of Ly$\alpha$ line EWs is consistent with a conversion factor of 70\%, as used in \citet{Hibon2010}, when assuming that the EW lower limits are the real values. Conversely, \citet{Kobayashi2010} predict a distribution of Ly$\alpha$ EWs clearly peaked toward high values, hence favouring conversion factors closer to 100\%. In the following, we therefore use these two values (70\% and 100\%) when converting \textit{NB1060} magnitudes to line luminosities.

\begin{figure}[bt]
\centering
\resizebox{\hsize}{!}{\includegraphics{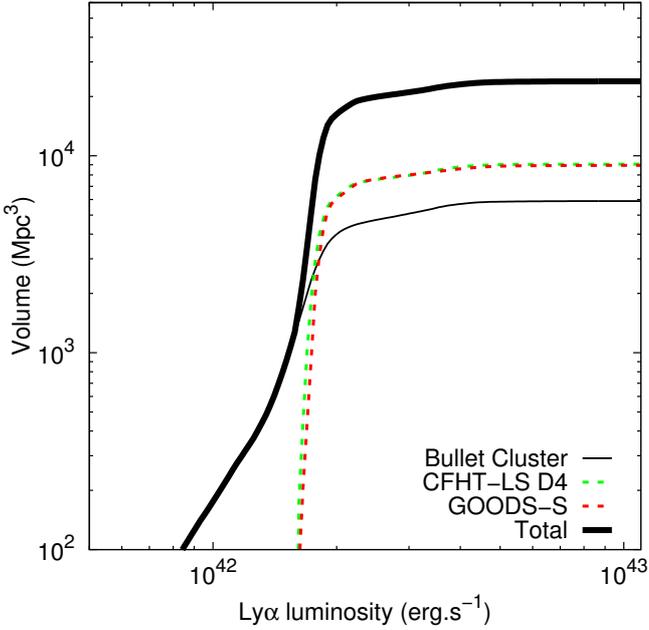}}
\caption{The comoving volume $V(L)$ sampled by the \textit{NB1060} images as a function of Ly$\alpha$ luminosity. The two dashed curves correspond to the two blank fields (green -- CFHTLS-D4; red -- GOODS-S). The thin black curve corresponds to the Bullet Cluster. The thick black curve is the total comoving volume corresponding to the three fields. The Ly$\alpha$ luminosity corresponds to a 5$\sigma$ \textit{NB1060} limiting magnitude, assuming a 70\% average conversion factor between the Ly$\alpha$ luminosities and \textit{NB1060} magnitudes (see text).}
\label{fig:volume}
\end{figure}

In total, the comoving volume sampled by our images is $5.9\times 10^3~\mathrm{Mpc^3}$ for the Bullet Cluster and $9\times 10^3~\mathrm{Mpc^3}$ for each of the two blank fields, corresponding to a grand total of $\sim2.4\times 10^4~\mathrm{Mpc^3}$ for the three fields. The volume-luminosity relation is shown in figure~\ref{fig:volume}, and the effect of the gravitational amplification for the Bullet Cluster is clearly visible, where the gravitational amplification enables probing fainter luminosities, however over increasingly smaller volumes.

\subsection{Cosmic variance and Poisson noise} \label{sec:cv}
We do not detect any LAE candidates down to a \textit{NB1060} 5$\sigma$ limiting magnitude of $\sim$25.9 to $\sim$26.1. To constrain the luminosity function of $z = 7.7$ Ly$\alpha$ emitters and compare our results with others, we make use of the Schechter formalism \citep{Schechter}:
\begin{equation}
\phi(L)\mathrm{d}L=\phi^{*}\left(\frac{L}{L^{*}}\right)^{\alpha}\mathrm{exp}\left(-\frac{L}{L^{*}}\right)\mathrm{d}\left(\frac{L}{L^{*}}\right)
\end{equation}
where $L^*$ is the characteristic luminosity defining the LF high luminosity cutoff, $\phi^*$ a volume density normalization factor, and $\alpha$ the faint-end slope characterizing how steeply the LF increases at low luminosities. Limited samples and large errors lead to degneracy between these three LF parameters. Pending more observational data and more accurate determinations of the high-z LAE LF parameters, most of the authors in the literature have settled on a canonical faint end slope value of $\alpha = -1.5$. Our observations probe luminosities similar, or slightly fainter, than those observed by other groups; therefore, we can compare our results to others, and to this aim we similarly adopt, unless stated otherwise, a faint end slope of $\alpha$ = -1.5.

Denoting $V(L)$ the comoving volume probed by our observations as a function of the luminosity $L$, as described in section~\ref{sec:volume} and shown in figure~\ref{fig:volume}, the total number of objects $N(L^*,\phi^*,\alpha)$ is given by
\begin{equation}
N(L^*,\phi^*,\alpha)=\int \phi(L)V(L)CF(L)\mathrm{d}L
\label{eq:number_counts}
\end{equation}
where $CF(L)$ is the completeness function (see section~\ref{sec:completeness}). The conversion of the completeness function as a function of magnitude as shown in figure~\ref{fig:completeness} to $CF(L)$ takes into account the conversion factor of 70\% and 100\% mentioned above, and in the case of the Bullet Cluster further takes the amplification map of the cluster into account.

From a Poisson distribution, one can easily compute single-sided confidence levels (CL) for the upper limits of the expected number of events $N_u$ that correspond to a measured number of events $n$, as \citep{Gehrels1986}:
\begin{equation}
\sum_{k=0}^{n}\mathrm{e}^{-N_u}\frac{N_u^k}{k!}=1-\mathrm{CL}\,.
\end{equation}

In our situation of zero detection ($n=0$), the 84.13 , 97.72, 99.87, and 99.99 percentiles\footnote{\label{note_gauss_sigma}Corresponding to the single sided 1$\sigma$,2$\sigma$,3$\sigma$,4$\sigma$ percentiles of a normal distribution} confidence levels correspond to upper limits of the mean number of events $N_u$ of 1.84, 3.78, 6.61, and 10.36, respectively. The situation $N_u \le 1$ corresponds to a 63\% confidence level. Therefore, with zero detection and assuming pure Poisson statistics, one can exclude, at a given confidence level CL, the  luminosity function parameters that would yield an expected number of objects $N_u$ with our survey parameters.


However, considering the somewhat limited area covered by our observations, we need to consider the effects of cosmic variance in our statistical analysis. \citet{Somerville2004} are among the first authors to derive quantitative estimates of the effects of cosmic variance from cold dark matter (CDM) models and observations of the two-point correlation functions of galaxy populations in the GOODS survey data. \citet{Trenti2008} expand on this work and propose a cosmic variance model, based on N-body simulations and halo occupation distribution models, and applied to a variety of high-z galaxy populations. We use below the on-line version of this model to estimate the effects of cosmic variance on our observations.

Various prescriptions have been proposed for the distribution function of galaxy number counts affected by cosmic variance, see \citet{Yang2011} for a recent discussion and analysis of the galaxy counts-in-cells distribution functions in the SDSS data. Although not physically motivated, the negative binomial distribution (NBD) fits the SDSS data well at both low and high number counts and provides a convenient description for the distribution function of galaxies with positive number counts and overdispersion relative to a Poisson distribution. To prevent technical problems with the use of normal or lognormal distributions (e.g. truncation to positive numbers), we chose to adopt the NBD as an ad hoc representation of the probability density function of low galaxy number counts. The NBD can be conveniently expressed as a Poisson random variable whose mean population parameter is itself random and distributed as a Gamma distribution of variance equal to the relative cosmic variance.

By definition of the cosmic variance, the variance $\overline{N^2} - \overline{N}^2$ of the number of galaxies of mean number $\overline{N}$ is in excess of the Poisson variance $\overline{N}$ and is given by
\begin{equation}
\overline{N^2} - \overline{N}^2 = \overline{N}^2 \times \sigma_v^2 + \overline{N}
\end{equation}
where $\sigma_v^2$ is the relative cosmic variance.
With the NBD prescription, one can derive the confidence level CL corresponding to no detections in our observations, for a known $\sigma_v^2$. To estimate the cosmic variance, we run the on-line cosmic variance calculator\footnote{http://casa.colorado.edu/$\sim$trenti/CosmicVariance.html} \citep{Trenti2008} and derive $\sigma_v^2$ for low number counts $\overline{N}$. We use the parameters corresponding to our observations: the instrument's field of view, a mean redshift of 7.74, and a redshift interval of 0.08. We select the Sheth-Tormen bias formalism, a value of 0.8 for $\sigma_8$, and a value of 1.0 for the user-selectable halo-filling factor. Although the filling factor of LAEs is likely to be significantly lower than 1.0, see e.g. \citet{Ouchi2010}, adopting a value of 1.0 leads to overestimating the importance of cosmic variance and therefore corresponds to a worst-case scenario in our analysis. Finally, we define a completeness factor of 1.0 in the cosmic variance calculator since we use our measured completeness function in equation~\ref{eq:number_counts} to derive the number of galaxies that correspond to a given set of LF parameters. We then fit the resulting relative cosmic variance $\sigma_v^2$ corresponding to our survey parameters as a function of $\overline{N}$, and we plug this cosmic variance into a Gamma distribution. For each pair of ($L^*,\phi^*$) parameters we thus derive a confidence level corresponding to no detections in our observations. 

\begin{figure}[t]
\centering
\resizebox{\hsize}{!}{\includegraphics{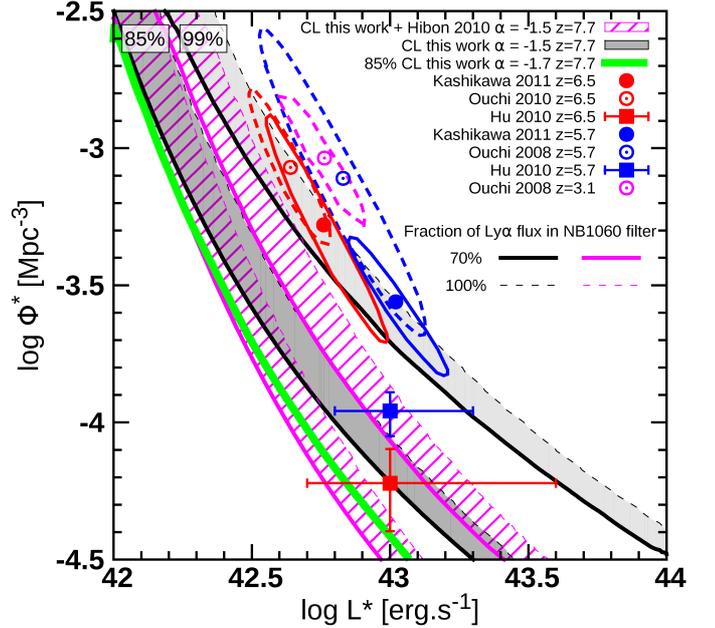}}
\caption{Parameters of the $z = 7.7$ luminosity function excluded at 85\% and 99\% confidence levels from our data, assuming the Schechter formalism and a fixed faint end slope $\alpha=-1.5$ unless otherwise stated. Filled circles correspond to the best fit LF parameters at $z=5.7$ (blue) and $z=6.5$ (red) from \citet{Kashikawa2011}. The ellipses correspond to the 3$\sigma$ confidence levels for these datapoints. Open circles correspond to the best fit LF parameters of the photometric samples of \citet{Ouchi2008} at $z=3.1$ (magenta) and $z=5.7$ (blue) and of \citet{Ouchi2010} at $z=6.5$ (red). The ellipses correspond to the 2$\sigma$ confidence levels for these datapoints. The filled square symbols correspond to the best fit LAE LF parameters at $z=5.7$ (blue) and $z=6.5$ (red) from \citet{Hu2010}. The plain (resp. dotted) black lines correspond to the 85\% and 99\% confidence levels corresponding to no detections in our  HAWK-I data (this work), assuming a conversion factor of 70\% (resp. 100\%) between the \textit{NB1060} and Ly$\alpha$ fluxes. The dark (resp. light) grey zones correspond to the range of 85\% (resp. 99\%) confidence levels for conversion factors between 70\% and 100\%, as delimited by the plain and dotted lines. Similarly, the lines and hatched zones in magenta colour correspond to the 85\% and 99\% confidence levels when adding to our work the null spectroscopic confirmation of the five brightest $z = 7.7$ LAE candidates of \citet{Hibon2010}.}
\label{fig:lf_param_limit}
\end{figure}

This is represented in figure~\ref{fig:lf_param_limit} together with the best-fit parameters of the Ly$\alpha$ LAE luminosity functions at redshifts 5.7 and 6.5 from various sources. We report in this plot the range of  ($L^*,\phi^*$) parameters excluded at 85\% and 99\% confidence levels, for conversion factors between \textit{NB1060} magnitudes and Ly$\alpha$ luminosities varying from 70\% and 100\% (see discussion in section~\ref{sec:volume}). We report the upper ($L^*,\phi^*$) exclusion zones derived from the absence of detections in the HAWK-I observations only (this work). Furthermore, we report the upper exclusion zones at the same confidence levels when adding to the null HAWK-I detections the null spectroscopic confirmation of the five brightest $z=7.7$ LAE candidates\footnote{\label{note_hibon}Indeed, spectroscopic follow-up with X-shooter at the VLT of the five brightest candidates failed to detect Ly$\alpha$ emission. These results will be reported separately in Cuby et al. (in prep.).} presented in \citet{Hibon2010}. Finally, for illustration, we also report the 85\% confidence level derived from this work (HAWK-I observations only) for a faint end slope of the $z=7.7$ Ly$\alpha$ LAE LF $\alpha=-1.7$. 

\section{Discussion}
\label{sec:discussion}
The various LF parameters reported in the literature at redshifts 5.7 and 6.5 present some significant differences. We report in figure~\ref{fig:lf_param_limit} some of the best-fit Schechter parameters at redshifts 3.1, 5.7 and 6.5. We report photometric samples partially confirmed in spectroscopy and purely spectroscopic samples. At the same redshift, the two series of datapoints clearly differ by significant amounts.

\citet{Ouchi2010}, and previously \citet{Kashikawa2006} followed by \citet{Kashikawa2011}, convincingly claim that the evolution, mostly in luminosity, of the $z\sim6.5$ LF from the lower redshift LFs at $z\sim3.1$ and $z\sim5.7$ LFs is a signature of reionization, due to a neutral-hydrogen fraction $x_\mathrm{HI}$ of the order of 20\% at $z = 6.5$. Conversely, the \citet{Hu2010} evolution of the LF parameters between the two same redshifts is mostly in density; accordingly, they do not infer a signature of reionization, in agreement with an earlier claim by \citet{Malhotra2004} that there was no evidence of reionization in the evolution of the LAE LF between these two redshifts.

In the light of the recent work from \citet{Hu2010}, the question of the evolution of the Ly$\alpha$ LAE luminosity function beyond redshift 5.7 becomes more difficult to comprehend. If the $z = 5.7$ point of \citet{Hu2010} \citep*[and][]{Malhotra2004} is correct, the picture by which the LAE LF would be approximately constant between $z = 3$ and $z = 5.7$, and would evolve significantly beyond redshift $\sim 6$ as a consequence of an increasingly neutral intergalactic medium (IGM), might need to be revisited. How useful are our results in this context?

From figure~\ref{fig:lf_param_limit}, we infer that we can safely exclude a non evolution of the LF parameters at $z = 7.7$ at more than 99\% confidence level from the \citet{Kashikawa2006,Kashikawa2011}, \citet{Ouchi2008} and \citet{Ouchi2010} values at $z = 5.7$ and 6.5. In that case, we should find $\sim13.7$ and $\sim11.7$ LAEs in the three HAWK-I fields based on the LF parameters estimated by \citet{Ouchi2010} and \citet{Kashikawa2011} at $z = 6.5$, respectively. The LF parameters of \citet{Hu2010} would predict $\sim2.5$ LAEs at $z = 6.5$ and can be similarly excluded at an $\sim 85\%$ confidence level, or higher, when including the null spectroscopic confirmation of the brightest LAE candidates of \citet{Hibon2010} (see previous section and footnote~\ref{note_hibon}). Our results therefore clearly show that the $z = 7.7$ Ly$\alpha$ LAE LF does evolve from lower redshifts, but in the absence of concordance between the data at lower redshifts, it is difficult to ascribe this evolution to the galaxy properties or to reionization.

\begin{figure}[ht]
\centering
\resizebox{\hsize}{!}{\includegraphics{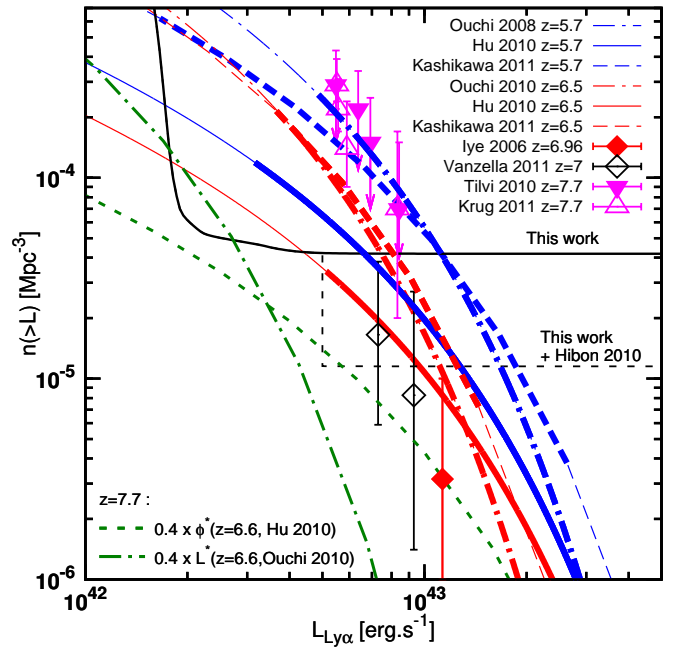}}
\caption{Cumulative Ly$\alpha$ luminosity functions. The blue (resp. red) lines show the cumulative luminosity functions at $z = 5.7$ (resp. $z = 6.5$) from \citet{Ouchi2008} (dotted-dashed lines), \citet{Hu2010} (plain lines) and \citet{Kashikawa2011} (dotted lines). The transitions to thin lines indicate the range of the luminosities probed by the observations. A fixed faint end slope of -1.5 is assumed. The parameter space probed by the HAWK-I observations is delimited by the black line, with the black dashed line corresponding to the HAWK-I observations and the null detection of the bright candidates of \citet{Hibon2010} mentioned in the text. The green dotted line corresponds to a 60\% evolution in density of the \citet{Hu2010} LF, and the short-dashed-long-dashed line corresponds to a 60\% evolution in luminosity of the \citet{Ouchi2010} $z = 6.5$ LF. The red filled diamond corresponds to the \citet{Iye2006} LAE detection at a redshift of 6.96. The black open diamonds correspond to the \citet{Vanzella2011} LAE detections at redshifts of 7.008 and 7.109. The magenta downward-pointing filled (resp. open) triangles are the photometric candidates at $z = 7.7$ from \citet{Tilvi2010} (resp. \citet{Krug2011}).}
\label{fig:lf_limit}

\end{figure}

To continue the discussion, we now consider two scenarios for the evolution of the LAE LF between $z = 6.5$ and $z = 7.7$, either in density or in luminosity. These scenarios are purely phenomenological and are not supported by theoretical considerations, only serving the purpose of assessing what the consequences of our results might be. This is illustrated in figure~\ref{fig:lf_limit} in a cumulative luminosity function plot. For a luminosity function to be consistent with our null results, it should lie, ignoring statistical fluctuations, below the contours corresponding to the parameter space probed by our observations. 

In the first scenario, we consider a $\sim$ 60\% evolution in density from the $z = 6.5$ datapoint of \citet{Hu2010}. Such an evolution can be entirely ascribed, in principle, to an intrinsic evolution of the LAE LF between $z = 6.5$ and $z = 7.7$ as in the models of \citet{Kobayashi2007} where the $z = 6.5$ LF, multipling $\phi^*$ by 0.4, almost perfectly coincides with the $z = 7.7$ LF. In practice, such an evolution could result from the combined evolution of the density of dark matter haloes and of the intrinsic galaxy properties. We also note that the UV LF of high-redshift galaxies between $z\sim4$ and $z \sim 8$ evolves essentially in luminosity \citep{Bouwens2011}, so a pure evolution in density should be treated with caution. As can be seen in Figs.~\ref{fig:lf_param_limit} and~\ref{fig:lf_limit}, such an evolution would fit most of the observational data, the shapes of the lower-redshift LFs, the datapoint from \citet{Iye2006} at $z = 6.96$, the datapoints from \citet{Vanzella2011} at $z=7.008$ and $z=7.109$, and our data. The mere existence of the \citet{Iye2006} and \citet{Vanzella2011} datapoints and of the \citet{Ota2010} LAE candidates at $z\sim7$ provides interesting constraints at high luminosities that cannot be fully captured in the ($L^*$, $\phi^*$) parameter space of figure~\ref{fig:lf_param_limit} and are better seen in figure~\ref{fig:lf_limit}. The main conclusion to be drawn from this test case is that to be consistent with our results, it does not require invoking a change in the Ly$\alpha$ IGM transmission and therefore a change in the neutral-hydrogen fraction of the IGM.

In the second scenario, conversely, we consider a 60\% change in luminosity from the  $z=6.5$ LF of \citet{Ouchi2010}. This scenario clearly requires, by construction, a significant quenching of the IGM Ly$\alpha$ transmission. The present luminosity evolution scenario may be compatible with the claim by \citet{Fontana2010} of a significant quenching of Ly$\alpha$ emission among LBGs at $z\sim7$. Although strongly model dependent, a high neutral-hydrogen fraction of the IGM due to a still incomplete reionization would then be required in this scenario, e.g. $x_\mathrm{HI} \sim 60\%$ according to the model of \citet{Santos2004}.

To conclude this discussion, we cannot safely decide on whether we are `seeing' signatures of reionization in our results, as this depends on which assumptions we use for the LFs at lower redshifts. We call for clarifications of the LAE luminosity functions at redshift 6.5 and below, a prerequisite to drawing firmer conclusions on reionization as inferred from LAE observations. We also note that our results are inconsistent with the photometric samples from  \citet{Tilvi2010} and \citet{Krug2011} at $z = 7.7$, but spectroscopic confirmation is required before drawing firmer conclusions. In the meantime, observations at $z = 7.7$ continue. An ideal complement to the HAWK-I observations presented in this paper would be to probe higher luminosities (of the order of $10^{43}$ ergs.s$^{-1}$) over significantly larger comoving volumes (a few times 10$^6$ Mpc$^3$).

%
%

\section{Conclusion}
We searched for Ly$\alpha$ emitters in three fields, two blank fields and one cluster field with the HAWK-I instrument at the VLT, using a NB filter centred at 1.06 $\mu$m. Our data in this filter total 80 hrs of integration time. The total comoving volume is $\sim2.4\times 10^4\mathrm{Mpc^3}$. We reached a 5$\sigma$ limiting magnitude that we use as detection threshold, of $\sim 26.0 \pm 0.1$ AB magnitude. We selected the objects from various colour criteria, which depend on the auxiliary data available for each field. We did not detect any object matching our selection criteria that would correspond to $z = 7.7$ LAEs. We modelled the probability density function of high-z LAE populations including Poisson statistics and cosmic variance with a negative binomial distribution. From this statistical description and from the absence of LAE detections, we exclude a non-evolution scenario at $\sim 85\%$ (resp. 99\%) confidence level from the $z = 6.5$ LF of \citet{Hu2010} (resp. \citet{Ouchi2010,Kashikawa2011}). The large differences between the published LFs at $z = 6.5$ prevent us from inferring a robust estimate of the Ly$\alpha$ IGM transmission at $z = 7.7$, and therefore of the neutral-hydrogen fraction $x_\mathrm{HI}$ at this redshift. However, in all cases but the \citet{Hu2010} LF, a significant quenching of the Ly$\alpha$ transmission by the IGM, probably due to reionization, is required.

\begin{acknowledgements}
We acknowledge financial support from Agence Nationale de la Recherche (grant ANR-09-BLAN-0234-01). FC is supported by the Swiss National Science Foundation (SNSF). The Dark Cosmology Centre is funded by the Danish National Research
Foundation.

\end{acknowledgements}

\bibliographystyle{aa}
\bibliography{hawki_LP_final.bib}

\end{document}